\documentclass[trackchanges,twocolumn]{aastex701}

\usepackage{amsmath}

\revised{\today}
\usepackage{appendix}
\usepackage{stfloats}
\usepackage{afterpage}

\usepackage{graphicx}
\usepackage{subcaption}
\usepackage{multirow}
\usepackage{txfonts}
\usepackage{lipsum}
\usepackage{subcaption}         
\usepackage{txfonts}         
\usepackage{color}
\usepackage{hyperref}
\usepackage{natbib}
\usepackage{multirow}                                 
\usepackage{placeins}           

\graphicspath{{./}{figure/}}

\begin{document}
\title{Effects of Ambient Medium Profiles on the Evolution of Relativistic Magnetized Jets}

\author[0009-0006-9884-6128]{Xu-Fan Hu}
\affiliation{Tsung-Dao Lee Institute, Shanghai Jiao-Tong University, Shanghai, 1 Lisuo Road, 201210, People's Republic of China}
\email{ebr105@sjtu.edu.cn}

\author[0000-0002-8131-6730]{Yosuke Mizuno}
\affiliation{Tsung-Dao Lee Institute, Shanghai Jiao-Tong University, Shanghai, 1 Lisuo Road, 201210, People's Republic of China}
\affiliation{School of Physics \& Astronomy, Shanghai Jiao-Tong University, Shanghai, 800 Dongchuan Road, 200240, People's Republic of China}
\affiliation{Key Laboratory for Particle Physics, Astrophysics and Cosmology (MOE), Shanghai Key Laboratory for Particle Physics and Cosmology, Shanghai Jiao-Tong University, 800 Dongchuan Road, Shanghai, 200240, China}
\email{mizuno@sjtu.edu.cn}

\correspondingauthor{Yosuke Mizuno}
\email{mizuno@sjtu.edu.cn}

\begin{abstract}
Relativistic radio-galaxy jets are conventionally divided into two morphological classes: FR~I and FR~II.
Among them, FR~I jets have relatively low radio luminosities and edge-darkened morphologies, indicating they undergo substantial deceleration and disruption on sub-kiloparsec to kiloparsec scales, often developing a turbulent and plume-like morphology.
Here, we investigate the evolution of relativistic magnetized jets in stratified ambient media using 2D and 3D relativistic magnetohydrodynamic simulations with several ambient profiles.
When the jet and ambient pressures are initially balanced, a Bondi-like atmosphere confines the jet to a broad parabolic shape ($R_j\propto z^{0.649\pm0.003}$) and produces a series of recollimation shocks.
Over-pressured jets instead produce a conical shape and allow the development of the current-driven kink instability in the downstream.
Based on the simulations, we derive a kink-stability criterion that depends primarily on the jet power and ambient pressure.
We also calculate multi-frequency emission maps using a special relativistic radiative transfer code.
Each recollimation shock appears as a bright knot, resembling the knot complex downstream of HST-1 of the M87 jet, whereas a jet with a single recollimation shock develops a bright feature close to its base and thus exhibits an FR~I-like morphology.

\end{abstract}


\section{Introduction}\label{intro}

Relativistic jets are among the most striking plasma phenomena in the universe. These highly collimated outflows can extend from subparsec scales into the circumgalactic medium \citep[e.g.,][]{blandford2019,oei2024}.

Radio jets are traditionally classified into two morphological classes \citep{fanaroff1974}.
FR~I jets are typically less luminous and edge-darkened: their emission is brightest near the nucleus and fades downstream (e.g., M87; \citealt{asada2012,hada2024}).
By contrast, FR~II jets are generally more luminous and edge-brightened, with compact terminal hotspots in which strong jet-ambient interactions produce termination shocks (e.g., Cygnus~A and 3C~295; \citealt{perley1984,harris2000}).
Recent studies \citep[e.g.,][]{ghisellini2011,sadler2014} have also shown that many local radio-loud active galactic nuclei (AGNs) exhibit little or no extended kiloparsec-scale emission \citep{baldi2023}.
These objects, commonly referred to as FR~0 sources, are often regarded as a low-power extension of the FR~I population \citep{costa2024}.

Although the magnetohydrodynamic (MHD) framework for jet formation was established decades ago \citep{blandford1977,blandford1982}, the physical origin of the diverse large-scale jet morphologies remains under debate.
Parameterized broadband spectral codes such as {\tt AGNJET} \citep{markoff2005,markoff2008} and {\tt JETSET} \citep{massaro2006,tramacere2009,tramacere2011} often prescribe a conical geometry and model jet acceleration, shock formation, and the resulting radiation semi-analytically \citep[e.g.,][]{rodi2021}.
Observations, however, indicate that the ambient medium strongly influences jet collimation \citep{asada2012}.
The analytical study of \cite{lyubarsky2009} identified a critical ambient-pressure index near $a=2$: jets remain parabolic for $a<2$ but become asymptotically conical for $a>2$.
Two-dimensional special relativistic MHD (SRMHD) simulations reached a similar conclusion \citep{komissarov2009}.

Axisymmetric models cannot capture three-dimensional, nonaxisymmetric instabilities.
For example, \cite{jortstad2022} reported signatures of a growing current-driven (CD) kink instability downstream of a recollimation feature during the 2020 outburst of BL~Lac.
The CD kink instability is expected to develop in Poynting-flux-dominated jets that carry large-scale helical magnetic fields \citep[e.g.,][]{moll2008,mizuno2009kink,mizuno2014}.
In addition, \cite{gomez2026} associated the twisted ridge-line structure of OJ~287 with the Kelvin-Helmholtz instability, which arises from velocity shear at the jet boundary.
Because many relevant instabilities are intrinsically three-dimensional \citep{gourgouliatos2018,abolmasov2023}, fully 3D SRMHD simulations are required.
Using such simulations, \cite{bromberg2016} showed that the external kink instability broadens the effective jet head and reduces its advance speed, and they proposed a stability criterion based on the ambient density and jet power.
Both factors have also been proposed as contributors to the FR dichotomy \citep{porth2015,tchekhovskoy2016}.
\cite{duran2017} considered broken ambient profiles and associated the transition radius with the Bondi scale, suggesting that changes in the ambient profile can localize dissipation in AGN jets and may produce features such as HST-1 in M~87 \citep[e.g.,][]{asada2012}.
Despite this progress, it remains unclear whether the ambient density or pressure is the principal parameter controlling the evolution of large-scale kink structures.

In this work, we perform 2D and 3D SRMHD simulations of propagating relativistic magnetized jets.
Previous studies have demonstrated the importance of jet power for the FR dichotomy \citep[e.g.,][]{porth2015,tchekhovskoy2016}.
In this work, we focus on the FR~I jets, which have relatively low jet power, and track the dynamical and energetic evolution of the jets across transitions between different ambient-medium profiles.
Here, we instead hold the jet power fixed and 
We then perform post-processing SRRT calculations and discuss the relationship between our results and observations of FR~I jets.

The paper is organized as follows.
Section~\ref{setup} describes the numerical methods and simulation setup.
Section~\ref{re} presents the jet dynamics, including recollimation shock and kink development.
Section~\ref{criterion} introduces the stability criterion derived from our simulations.
Section~\ref{rad} discusses the properties of the termination shocks and presents multi-frequency SRRT images.
Section~\ref{con} summarizes our conclusions.

\section{Numerical Setup}\label{setup}

We perform 2D and 3D simulations of propagating relativistic magnetized jets in Cartesian coordinates using the publicly available RMHD code {\tt PLUTO} \citep{mignone2007}.
We solve the RMHD equations in the following form:
\begin{align}
  \frac{\partial}{\partial t}(\gamma\rho)+\nabla\cdot(\gamma\rho \mathbf{v})=0, \label{1} \\
  \frac{\partial}{\partial t}(\omega_t\gamma^2\mathbf{v}-b^0\mathbf{b})+\nabla\cdot(\omega_t\gamma^2\mathbf{v}\mathbf{v}-\mathbf{b}\mathbf{b}+\mathbf{I} p_t)=\rho\mathbf{g}, \label{2} \\
  \frac{\partial}{\partial t}(\omega_t\gamma^2-b^0b^0-p_t)+\nabla\cdot(\omega_t\gamma^2\mathbf{v}-b^0\mathbf{b})=\mathbf{m\cdot g}, \label{3} \\
  \frac{\partial}{\partial t}\mathbf{B}+\nabla\cdot(\mathbf{v}\mathbf{B}-\mathbf{B}\mathbf{v})=0, \label{4}
\end{align}
Here, $b^0=\gamma\mathbf{v\cdot B}$, $\mathbf{b}=\mathbf{B}/\gamma+\gamma\mathbf{(v\cdot B)v}$, $\omega_t=\rho h+B^2/\gamma+\mathbf{(v\cdot B)^2}$, and $p_t=p_{gas}+[B^2/\gamma^2+\mathbf{(v\cdot B)^2}]/2$.
We set the speed of light to $c=1$.
The units of velocity $v$, density $\rho$, pressure $p$, and magnetic-field strength $B$ are $c$, $\rho_0$, $\rho_0c^2$, and $\sqrt{4\pi\rho_0c^2}$, respectively.

All jets are injected through a cylindrical nozzle at the bottom of the simulation domain with a relatively low initial Lorentz factor, $\gamma=2.6$, and a high magnetization, $\sigma$, as motivated by GRMHD simulations \citep[e.g.,][]{mckinney2006,nakamura2018}.
We assume the initial jets are axisymmetric and are set up in cylindrical coordinates.
We adopt an initial jet half-opening angle of $\theta=5^\circ$ \citep{mckinney2006}, which gives the cylindrical velocity components $v_z=v_0\cos\theta$, $v_R=v_0\sin\theta$, and $v_\phi=0$, where $v_0=\sqrt{1-1/\gamma^2}$.
The specific enthalpy is defined as $h=1+\Gamma p_{gas}/[(\Gamma-1)\rho]$.
The magnetization and plasma beta are defined as $\sigma=|\mathbf{b}|^2/\rho$ and $\beta=2p_g/b^2$, respectively. 
Here we note that the difference in jet half-opening angle does not affect the simulation results much. 
Therefore, we fixed one jet half-opening angle in this study.
We adopt an ideal equation of state with an adiabatic index of $\Gamma=4/3$.

\begin{figure}[!h]
  \centering
  \includegraphics[width=0.9\linewidth]{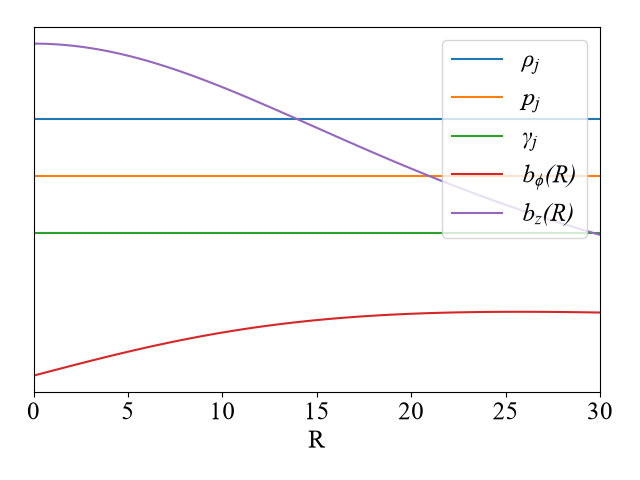}
  \caption{Initial radial profiles of the jet density, pressure, Lorentz factor, and fluid-frame magnetic field. The quantities $b_\phi(R)$ and $b_z(R)$ are shown to scale, whereas the other quantities are plotted in arbitrary units.}
  \label{fig:jet}
\end{figure}

The magnetic field is set only in the jet region, and the poloidal and toroidal components are prescribed as
\begin{equation}
  B_z=B_0\frac{PR_j}{A+R^2},
  B_\phi=B_0\frac{R}{A+R^2},\label{eq:mag}
\end{equation}
where $R_j=30$ is the initial jet radius, $R$ is distance from the jet axis, $P=(R/R_j)(B_z/B_\phi)$ is the magnetic pitch, and $A=\gamma^2P^2R_j^2$.
In this configuration, magnetic tension balances magnetic pressure, allowing the gas pressure to be uniform across the jet (see Appendix~\ref{mag} for the derivation of the magnetic profile).
We adopt an initial \textbf{magnetically dominated} jet parameters $\rho_j=10^3$, $p_j=60$, $P=0.33$, and $B_0=1.3\times10^4$, which yield $\langle\beta\rangle=0.005$ and $\langle\sigma\rangle=24$, consistent with \cite{duran2017}.
The initial jet fast magnetosonic Mach number along the z direction ${\cal M}_{fms,z}$ is around 0.63.
We check the snapshots at early simulation time, and do not see any discontinuity around the nozzle.
Figure~\ref{fig:jet} shows the radial profiles of the jet density, pressure, Lorentz factor, and fluid-frame magnetic field. 
We apply a sharp boundary between the jet and ambient medium.
We place a point-mass potential ($g$ term in equations~\ref{2} and \ref{3}) 
at the origin to represent the central black hole and maintain the stratification of the ambient medium. 
Thus, the ambient medium within the transition radius is in hydrostatic equilibrium.

\renewcommand{\arraystretch}{1.5}
\begin{table*}[!h]
  \caption{Basic properties of the simulation models.}
  \centering
  \begin{tabular}{lll}
    \hline \hline
    Model & Ambient density profile & Ambient pressure profile \\ \hline
    {\tt 2D}         & \multirow{2}{*}{$\displaystyle
      \begin{cases}
        10^{5}(r/10^2)^{-1.5} & r<10^3\\
        10^{3.5}(r/10^3)^{-1} & r>10^3
    \end{cases}$}            &  \multirow{2}{*}{$\displaystyle
      \begin{cases}
        8\times10^{3}(r/10^2)^{-2.5} & r<10^3\\
        8\times10^{0.5}(r/10^3)^{-1} & r>10^3
    \end{cases}$}\\
    {\tt 3D}       &                         & \\  \hline
    {\tt 3D0}      & $10^{5}(r/10^2)^{-1.5}$ & $8\times10^{3}(r/10^2)^{-2.5}$ \\ \hline
    {\tt 3Dl}       & $\displaystyle
    \begin{cases}
      10^{4}(r/10^2)^{-1.5} & r<10^3\\
      10^{2.5}(r/10^3)^{-1} & r>10^3
    \end{cases}$            &  $\displaystyle
    \begin{cases}
      8\times10^{2}(r/10^2)^{-2.5} & r<10^3\\
      8\times10^{-0.5}(r/10^3)^{-1} & r>10^3
    \end{cases}$ \\ \hline
    {\tt 3Dt}        & $\displaystyle
    \begin{cases}
      10^{4}(r/10^2)^{-1.5} & r<10^4\\
      10^{1}(r/10^4)^{-1} & r>10^4
    \end{cases}$            &  $\displaystyle
    \begin{cases}
      8\times10^{2}(r/10^2)^{-2.5} & r<10^4\\
      8\times10^{-3}(r/10^4)^{-1} & r>10^4
    \end{cases}$ \\
    \hline
  \end{tabular}
  \label{t1}
\end{table*}

\begin{figure}[!h]
  \centering
  \includegraphics[width=\linewidth]{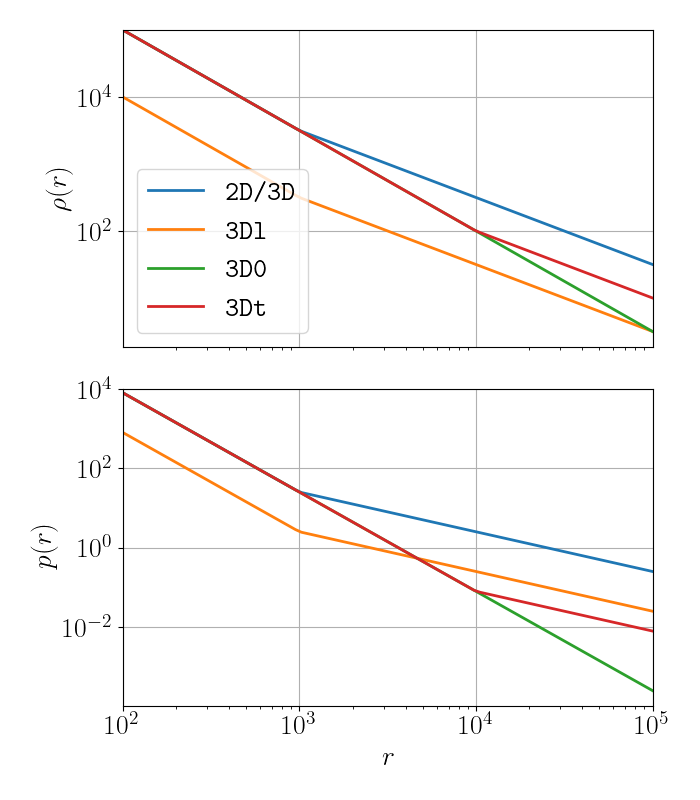}
  \caption{Density (upper panel) and gas-pressure (lower panel) profiles of the external medium for the different simulation models.}
  \label{fig:rho&prs}
\end{figure}

\begin{figure}[!h]
  \centering
  \includegraphics[width=\linewidth]{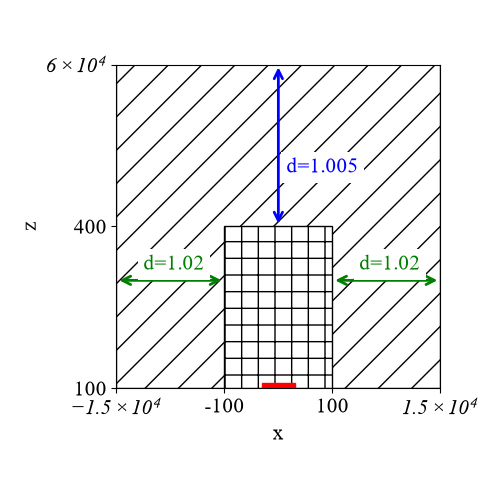}
  \caption{Schematic of the grid in the $x$-$z$ plane. The red block indicates the jet-injection nozzle ($|x|<30$), and the black rectangle encloses the uniform-grid region; stretched grids are used elsewhere. The dimensions are not shown to scale.}
  \label{fig:grid}
\end{figure}

We investigate the jet evolution for several radial profiles of the ambient medium. 
Here we assume the ambient medium is static and unmagnetized.
The inner ambient medium follows Bondi-like density and pressure profiles, $\rho_a\propto r^{-1.5}$ and $p_a\propto r^{-2.5}$, where $r$ is the radius in spherical coordinates.
The transition radius is set to $r=10^3$ in models {\tt 3D} and {\tt 3Dl}, and to $r=10^4$ in model {\tt 3Dt}.
Beyond this radius, we adopt a shallower density profile and a constant gas temperature, motivated by X-ray observations of the environment around M87 \citep{russell2015}.
We also perform an axisymmetric simulation (model {\tt 2D}) for comparison with model {\tt 3D}, which is performed in cylindrical coordinates.
In models {\tt 2D}, {\tt 3D}, and {\tt 3D0}, the jet \textbf{total} pressure is initially balanced with the ambient pressure.
Models {\tt 3Dl} and {\tt 3Dt} instead contain overpressured jets \textbf{with a pressure ratio of 10}, allowing \textbf{the initial jets to accelerate more effectively}.
Table~\ref{t1} summarizes the ambient density and pressure profiles of all models, and Figure~\ref{fig:rho&prs} shows these initial profiles.

The 3D computational domain covers $x,y\in[-1.5\times10^4,1.5\times10^4]$ and $z\in[10^2,6\times10^4]$.
To resolve the injected jet, we use a uniform grid with $100\times100\times50$ cells over $x,y\in[-100,100]$ and $z\in[100,400]$.
The injection nozzle occupies $R\in[0,30]$ and $z\in[100,110]$, so the jet radius at $z=100$ is $R_j=30$.
Stretched grids cover the remainder of the domain. In {\tt PLUTO}, a stretched grid satisfies
\begin{equation}
  \Delta x(d+d^2+\dots+d^N)=x_R-x_L,\label{eq2}
\end{equation}
where $\Delta x$ is the width of the nearest uniform cell, $d$ is the stretching ratio, $N$ is the number of cells in the stretched patch, and $x_L$ and $x_R$ are the patch boundaries.
In each transverse direction, we use 255 cells over both $[100,1.5\times10^4]$ and $[-1.5\times10^4,-100]$, corresponding to $d=1.02$.
In the longitudinal direction, we use 780 cells over $[400,6\times10^4]$, corresponding to $d=1.005$. 
The grid configuration is illustrated in Figure~\ref{fig:grid}.

We impose outflow conditions at all boundaries except the lower boundary at $z=10^2$.
At this boundary, the initial state is held fixed to enforce outward propagation of the jet and maintain stratified ambient medium structure.
This treatment is motivated by GRMHD simulations showing that both the jet and disk wind are launched by the central engine and flow outward within several hundred gravitational radii.
Here we injected the jet only, not the wind component.
We use linear spatial reconstruction and a second-order Runge-Kutta scheme for time integration.
For models {\tt 2D}, {\tt 3D}, and {\tt 3D0}, we employ divergence cleaning and the HLLC approximate Riemann solver.
For the initially unbalanced models {\tt 3Dl} and {\tt 3Dt}, we use the eight-wave formulation and the TVDLF Riemann solver.

\section{Results}\label{re}
\subsection{Recollimation of the 2D Jet}\label{2d}

\begin{figure}[!h]
  \centering
  \includegraphics[width=0.99\linewidth]{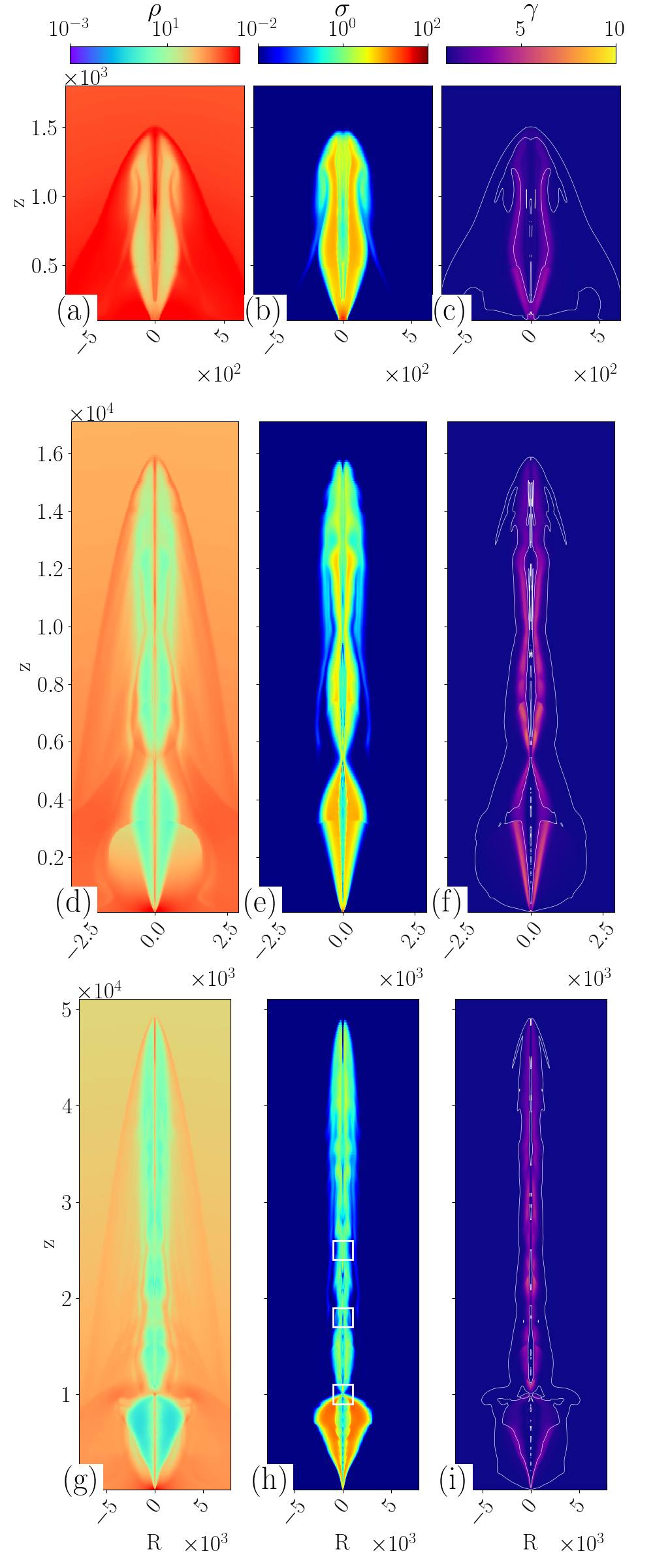}
  \caption{Axial maps of density (left column), magnetization (middle column), and Lorentz factor (right column) for model {\tt 2D}. The rows show snapshots at $t=2\times10^3$ (top), $2\times10^4$ (middle), and $6\times10^4$ (bottom). White boxes in panel (h) mark the recollimation shock regions. White contours in the right columns mark the regions where ${\cal M}_{fms,z}>1$.}
  \label{fig:2d}
\end{figure}

As discussed in Section~\ref{intro}, nonaxisymmetric MHD instabilities are suppressed in the 2D simulation, and the jet therefore remains stable as it propagates.
Figure~\ref{fig:2d} shows axial maps of the density, magnetization, and Lorentz factor for model {\tt 2D} at $t=2\times10^3$, $2\times10^4$, and $6\times10^4$.
Initially, the jet expands within the steep inner atmosphere and recollimates after crossing the transition radius (Figure~\ref{fig:2d}(a)-(c)).
Downstream of the recollimation shock, the magnetization decreases while the Lorentz factor increases, indicating efficient magnetic acceleration.
As the jet continues to propagate, the recollimation shock moves downstream (Figure~\ref{fig:2d}(d)-(f)).
A discontinuity is seen at about $z=3000$.
The downstream from $z=3000$  is compressed by the ambient medium and begins to recollimate, while at the upstream region ($z<3000$) jet accelerates in the environment where the ambient medium has been expelled.
Meanwhile, the jet at $z<3000$ entrains low-velocity ambient medium, resulting in the wide wing in Figure~\ref{fig:2d}(d).
We annotate the locations of the recollimation shocks with white boxes in panel (h).
The jet head retains a conical shape and does not develop a termination shock (Figure~\ref{fig:2d}(g)-(i)).

We next examine energy conversion along the jet. 
From the RMHD energy-momentum tensor, the magnetic energy flux is defined as
\begin{equation}
  \dot E_{mag}=\int\left[\gamma^2 v^z\left(\frac{B^2}{\gamma^2}+(\mathbf{v\cdot B})^2\right)-\gamma(\mathbf{v\cdot B})\left(\frac{B^z}{\gamma}+\gamma (\mathbf{v\cdot B})v^z\right)\right]dA\label{eq3}
\end{equation}
The kinetic energy flux is
\begin{equation}
  \dot E_{kin}=\int\gamma(\gamma-1) v^z\rho dA\label{eq4}
\end{equation}
and the internal energy flux is
\begin{equation}
  \dot E_{int}=\int\gamma^2 v^z\frac{\Gamma}{\Gamma-1}p_g dA\label{eq5}
\end{equation}
We evaluate these diagnostics only for material associated with the jet, identified using a passive tracer $Q$ injected through the nozzle.
Cells with tracer values above the adopted threshold are classified as jet material.
For example, the azimuthally averaged four-velocity of the jet material is calculated as
\begin{equation}
  \langle \gamma \beta \rangle=\frac{\int I \rho \gamma \beta\ dA}{\int I \rho dA}\label{eq6}\qquad
  I=
  \begin{cases}
    1 & Q>0.9\\
    0 & Q\le0.9
  \end{cases}
\end{equation}
where $I$ is an indicative function that includes the jet only.

\begin{figure}[!h]
  \centering
  \includegraphics[width=0.9\linewidth]{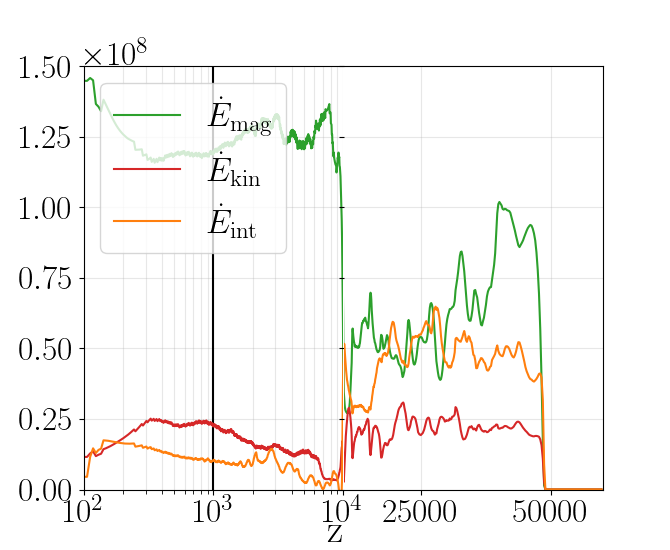}
  \caption{Magnetic (green), kinetic (red), and internal (orange) energy fluxes at $t=6\times10^4$.
  The black vertical line marks the transition radius.}
  \label{fig:2d_en}
\end{figure}

Figure~\ref{fig:2d_en} shows the azimuthally integrated magnetic, kinetic, and internal energy fluxes along the jet.
The initially Poynting-flux-dominated jet accelerates inside the transition radius, consistent with the observed acceleration and collimation zone (ACZ).
Because the transition radius is relatively small, the jet retains a substantial fraction of its magnetic energy after entering the flatter outer atmosphere.
At the recollimation shocks, some kinetic and internal energy is converted back into magnetic energy, producing enhancements near $z\sim2.5\times10^4$ and $z\sim3.7\times10^4$.
Downstream of these shocks, the magnetic energy is rapidly converted into kinetic and internal energy once again.

\subsection{Recollimation in 3D Jets}\label{3d}

\begin{figure}[!h]
  \centering
  \includegraphics[width=0.99\linewidth]{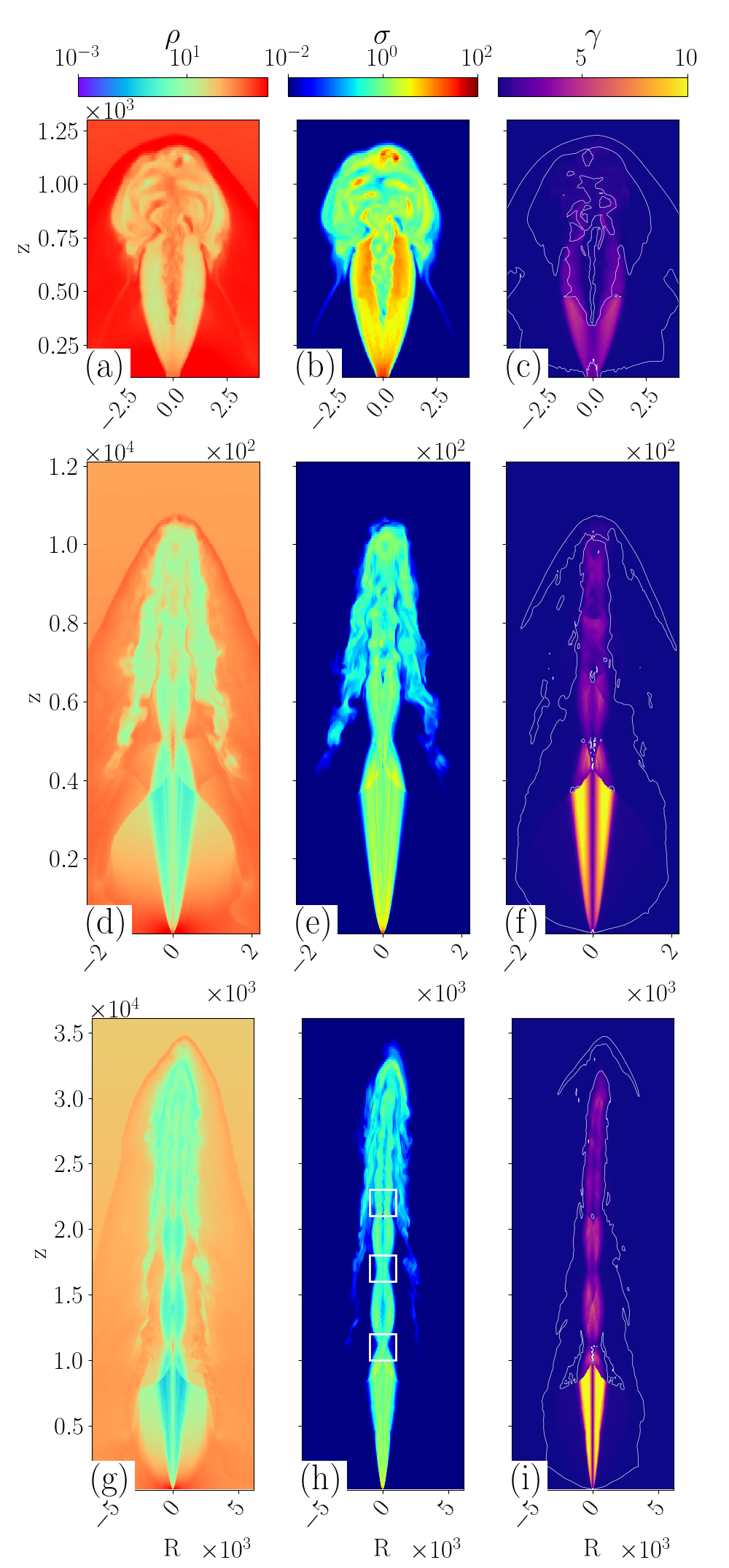}
  \caption{Same as Figure~\ref{fig:2d}, but for model {\tt 3D}. The recollimation points in panel (h) are $1.1\times10^4,1.7\times10^4,2.2\times10^4$.}
  \label{fig:3d}
\end{figure}

\begin{figure}[!h]
  \centering
  \includegraphics[width=0.85\linewidth]{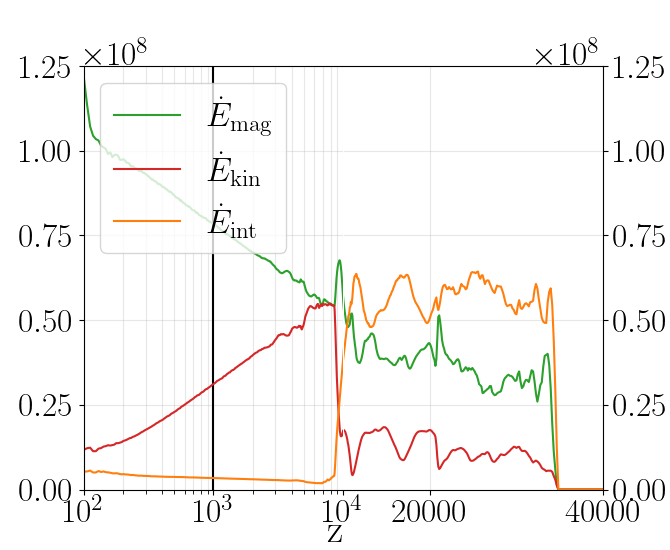}
  \includegraphics[width=0.99\linewidth]{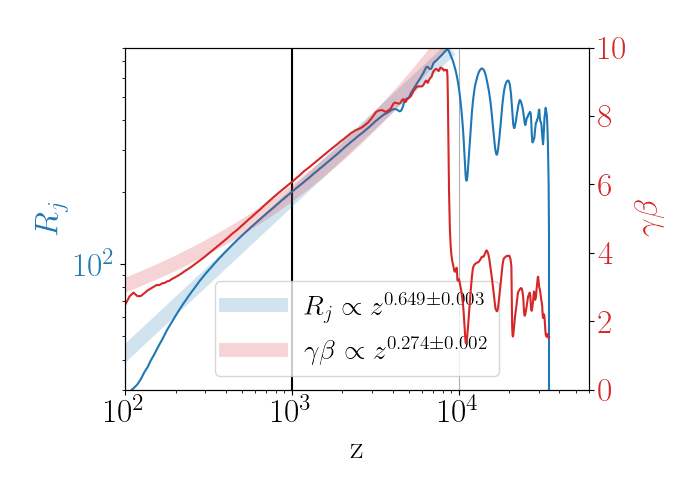}
  \caption{Top: same as Figure~\ref{fig:2d_en}, but for model {\tt 3D}. Bottom: azimuthally averaged jet radius $R_j$ (blue curve) and four-velocity $\gamma\beta$ (red curve) at $t=6\times10^4$. The corresponding broad, faint curves show the best-fit power laws.}
  \label{fig:3d_cut}
\end{figure}

Compared with the axisymmetric model {\tt 2D}, the 3D jet develops a more distorted head, indicating stronger interaction with the ambient medium (Figure~\ref{fig:3d}(a)-(c)).
Although the jet passes the transition raidus at $z>10^3$, its recollimation point forms farther downstream and continues to advance (Figure~\ref{fig:3d}(d)-(i)).
During this evolution, magnetic energy is efficiently converted into bulk kinetic energy (top panel of Figure~\ref{fig:3d_cut}).
Notably, the recollimation structures are much weaker and less regular in 3D model because non-axisymmetric modes distort the jet cross-section, mix jet and cocoon material, and redistribute the magnetic stress.
The bow-like contour appears in front of the jet head in Figure~\ref{fig:3d}(f,i), demonstrating the existence of shocked external gas (forward shock).

The bottom panel of Figure~\ref{fig:3d_cut} shows the azimuthally averaged jet radius and four-velocity.
The jet follows a broad parabolic profile, $R_j\propto z^{0.649\pm0.003}$, while the four-velocity is also well described by a power-law.
Its fitted index, $0.274\pm0.002$, is larger than the value of 0.16 measured for the M87 jet.

\begin{figure}[!h]
  \centering
  \includegraphics[width=0.99\linewidth]{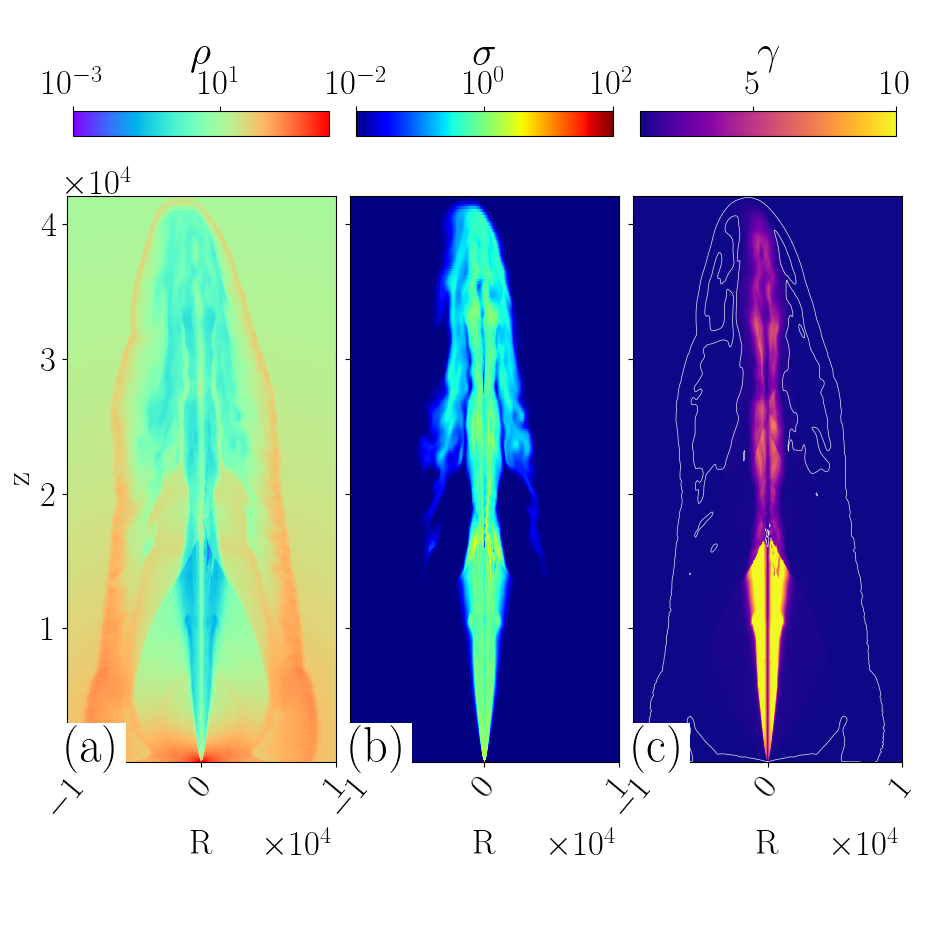}
  \caption{Same as Figure~\ref{fig:3d}, but for model {\tt 3D0} at $t=6\times10^4$.}
  \label{fig:3d0}
\end{figure}

After the jet passes the first recollimation point ($z>10^4$ in Figures~\ref{fig:3d}(g)-(i)), its four-velocity decreases sharply.
A similar deceleration has been observed downstream of HST-1 \citep{park2019}.
The flow downstream of the recollimation point becomes dominated by internal energy and develops a series of recollimation shocks without an obvious development of CD kink instability.

In model {\tt 3D0}, the jet retains a broad parabolic shape even in the absence of a transition radius, but it develops fewer recollimation shocks.
This result supports an association between the formation of multiple recollimation shocks and a flatter outer atmosphere, as suggested by \cite{russell2015}.

\subsection{Kink Instability in Overpressured Jets}\label{kink}

\begin{figure}[!h]
  \centering
  \includegraphics[width=0.99\linewidth]{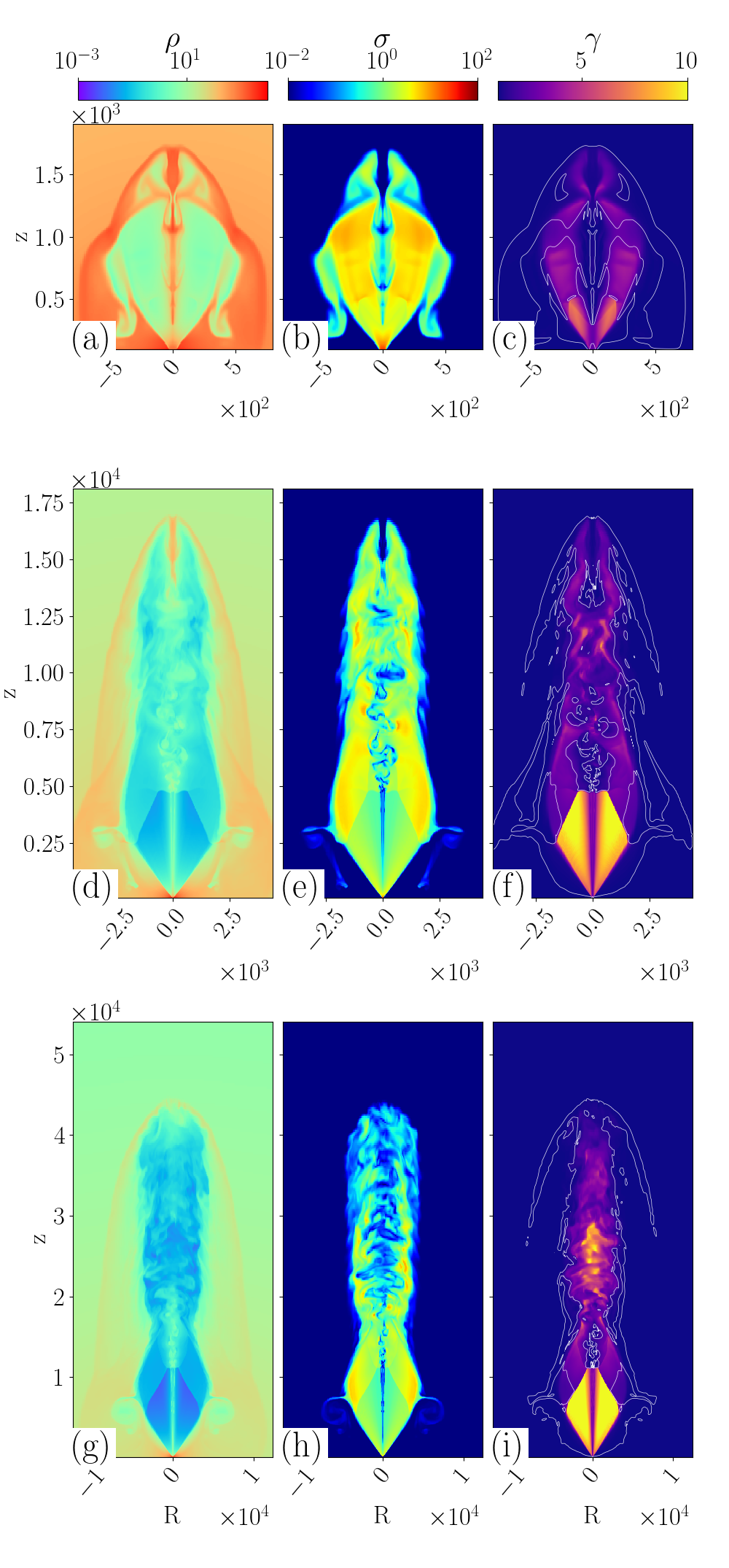}
  \caption{Same as Figure~\ref{fig:2d}, but for model {\tt 3Dl}.}
  \label{fig:3dl}
\end{figure}

\begin{figure}[!h]
  \centering
  \includegraphics[width=0.85\linewidth]{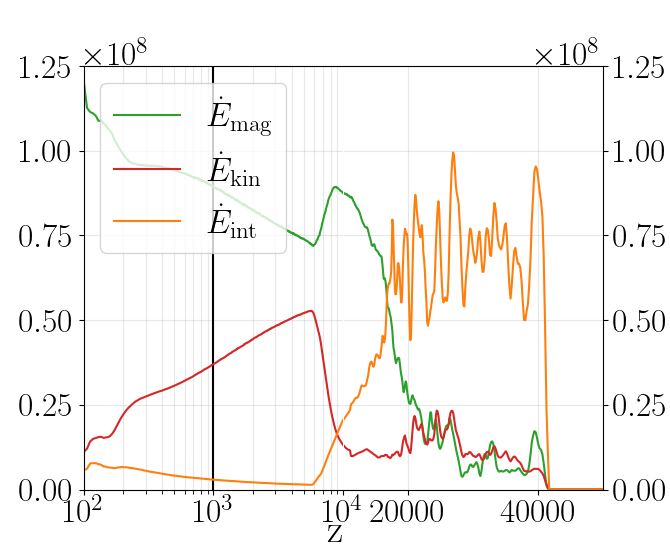}
  \caption{Same as Figure~\ref{fig:2d_en}, but for model {\tt 3Dl}.}
  \label{fig:3dl_en}
\end{figure}

The jet in model {\tt 3Dl} is initially over-pressured and therefore expands rapidly in the transverse direction, forming a conical profile during the early evolution (Figure~\ref{fig:3dl}(a)-(c)).
Consequently, its recollimated section is wider than those in models {\tt 3D} and {\tt 3D0}.
Its weaker interaction with the ambient medium also results in less dissipation.
Figure~\ref{fig:3dl_en} shows that the jet retains substantial magnetic energy in the downstream of the recollimation shock, enabling the growth of CD kink instability in the downstream flow (Figure~\ref{fig:3dl}(d)-(i)).

\begin{figure}[!h]
  \centering
  \includegraphics[width=\linewidth]{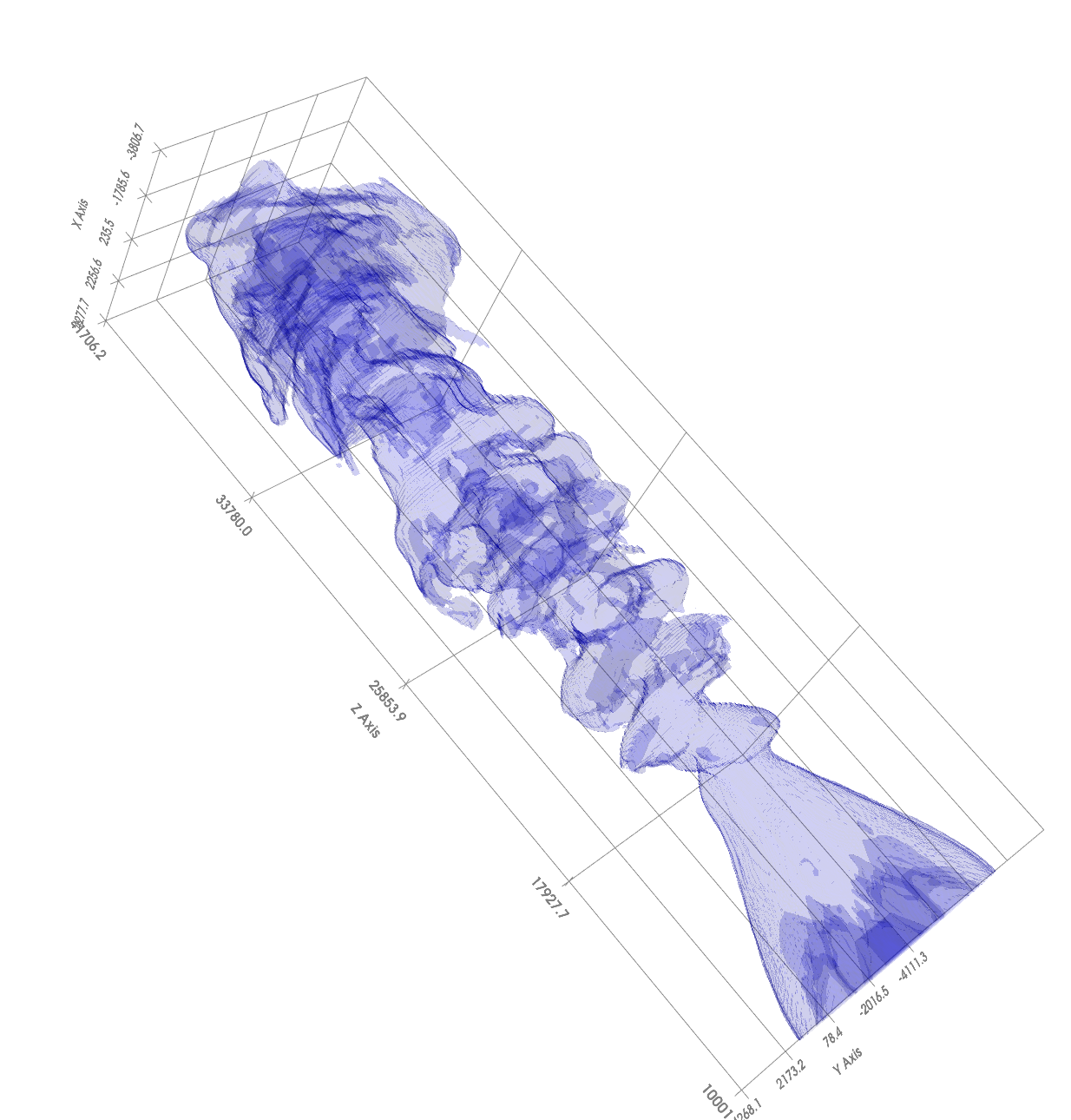}
  \caption{Three-dimensional density isosurface at $\rho_j=0.05$ for model {\tt 3Dl}.}
  \label{fig:kink}
\end{figure}

Notably, the developed CD kink instability disrupts only the inner jet region.
Figure~\ref{fig:kink} shows a three-dimensional isosurface density which indicates the twisted inner structure.
The surrounding jet layers shield the distorted jet spine from direct interaction with the ambient medium, allowing the downstream flow to remain relativistic.

\begin{figure}[!h]
  \centering
  \includegraphics[width=\linewidth]{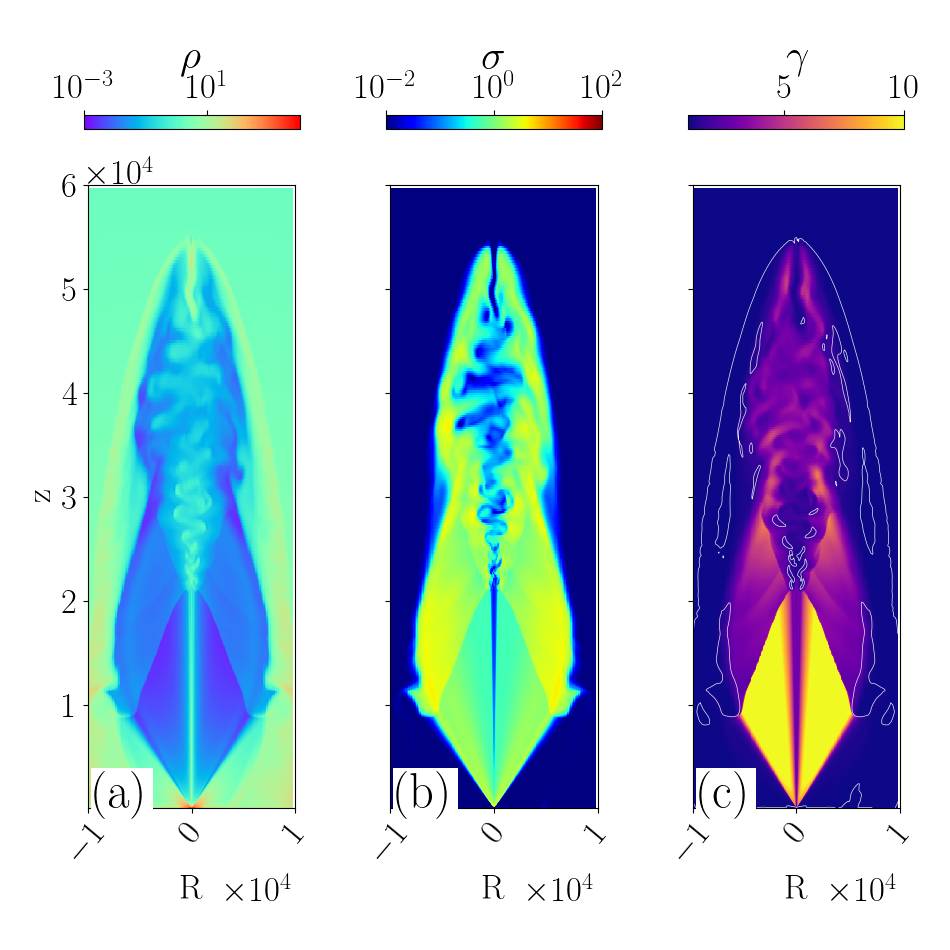}
  \caption{Same as Figure~\ref{fig:3d}, but for model {\tt 3Dt} at $t=6\times10^4$.}
  \label{fig:3dt}
\end{figure}

For model {\tt 3Dt}, we place the transition radius farther downstream, at $z=10^4$.
As shown in Figure~\ref{fig:3dt}, the jet expands to a larger radius and propagates farther than in model {\tt 3Dl}, but its qualitative evolution remains unchanged.

\subsection{Kink-stability Parameter}\label{criterion}

\begin{figure*}[!h]
  \centering
  \includegraphics[width=0.99\linewidth]{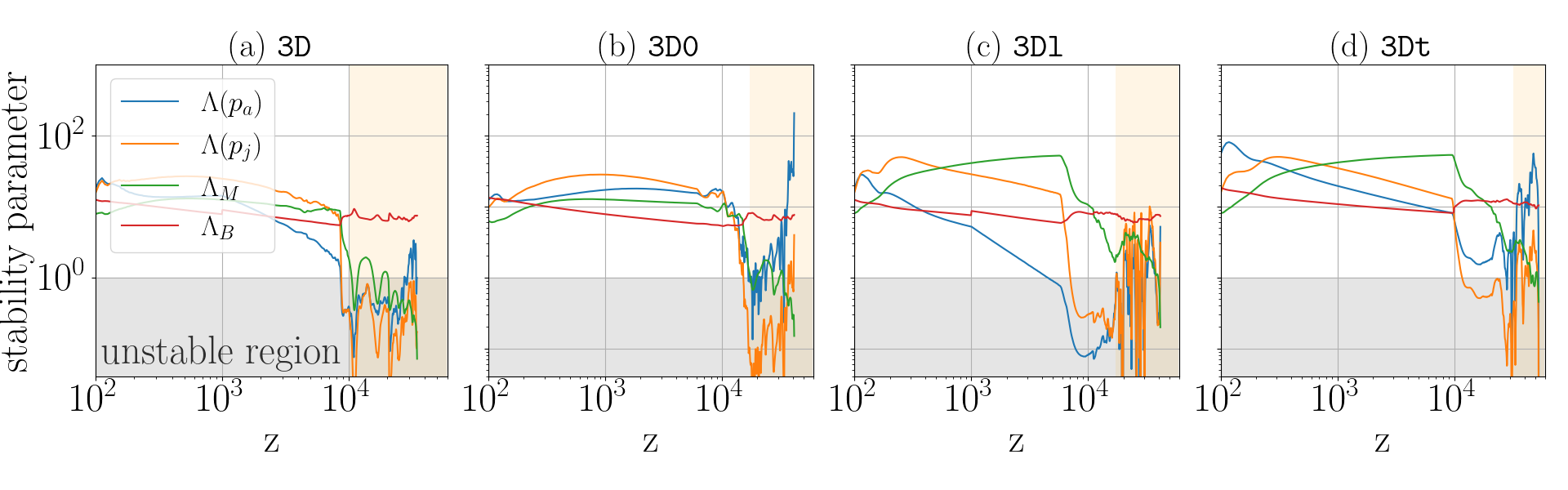}
  \caption{Comparison of the kink-stability parameters. The blue and orange curves show our parameters derived from the ambient pressure, $\Lambda(p_a)$, and jet pressure, $\Lambda(p_j)$, respectively. The green curve shows $\Lambda_M$ from \cite{mizuno2009kink}, and the red curve shows $\Lambda_B$ from \cite{bromberg2016}. The gray shaded region at the bottom indicates the predicted CD-kink-unstable regime, and the orange shaded region on the right of each panel indicates the internal-energy-dominated portion of the jet.}
  \label{fig:para}
\end{figure*}

To determine where the CD kink instability begins to develop, we first evaluate the stability parameter proposed by \cite{bromberg2016}:
\begin{equation}
  \Lambda_B=20\left(\frac{2\pi}{9}\right)^{1/2}\left[\frac{\pi(5-a)(3-a)}{6}\right]^{1/3}\left(\frac{L_j}{nm_p z^2 \gamma^2c^3}\right)^{1/6},\label{eq7}
\end{equation}
Here, $n$ is the number density of the ambient medium, $m_p$ is the proton mass, $z$ is the distance along the jet, and $a=-d\log n/d\log r$ is the logarithmic slope of the ambient density profile.
The criterion predicts that the jet is unstable to the CD kink mode when $\Lambda_B<1$.
However, it assumes equipartition between the toroidal and poloidal magnetic fields at the jet edge, an assumption that does not hold in our models.
As Figure~\ref{fig:para} shows, $\Lambda_B$ is inconsistent with our simulation results.
We therefore derive a new stability parameter appropriate for our magnetic-field configuration.

\cite{appl2000} expressed the distance traveled during one e-folding time as
\begin{equation}
  \frac{z}{R_j}\simeq7.52  \frac{b_z}{b_\phi}\frac{\gamma v}{v_A},\label{eq8}
\end{equation}
where ${b_z}/{b_\phi}$ is evaluated at the jet edge. Because $\sigma\gtrsim1$ and $\beta\lesssim1$ in our models, $v_A\sim c$, and the criterion can be written as
\begin{equation}
  \Lambda=7.52\frac{\gamma R_j}{z}\frac{b_z}{b_\phi}
\end{equation}
To express this criterion in terms of global jet properties, we begin with the jet power in code units:
\begin{align}
  m^{zt}&=\gamma^2 v^z w_t-b^0b^z \notag \\
  &=\gamma^2 v^z\left(\rho+\frac{\Gamma}{\Gamma-1}p_g+b^2\right)-\gamma(\mathbf{v\cdot B})\left[\frac{B^z}{\gamma}+\gamma (\mathbf{v\cdot B})v^z\right] \notag \\
  &\simeq \gamma^2 v^z \left(\rho+\frac{\Gamma}{\Gamma-1}p_g+b^2\right) \label{eq:energy}
\end{align}
where $w_t=\rho+\frac{\Gamma}{\Gamma-1}p_g+b^2$. 
The total jet power can then be approximated as
\begin{align}
  P_j&=\int m^{zt}\ 2\pi R c\ dR\approx \gamma^2 \left<  \rho+\frac{\Gamma}{\Gamma-1}p_g+b^2\right> \pi cR_j^2 \notag \\
  &\approx 2\pi\gamma^2R_j^2 c \langle p_j\rangle, \label{eq10}
\end{align}
where we used $\sigma>1$ and defined $p_j=p_g+b^2/2$. It follows that
\begin{equation}
  \gamma R_j=\sqrt{\frac{P_j}{2\pi c\langle p_j \rangle }}\label{eq:rj}
\end{equation}
The resulting stability parameter is
\begin{equation}
  \Lambda(p_j)=\frac{7.52}{z}\sqrt{\frac{P_j}{2\pi c \langle p_j \rangle}}\frac{b_z}{b_\phi}. \label{eq11}
\end{equation}

We next consider radial force balance (see Appendix~A of \citealt{hu2025aa}):
\begin{equation}
  \frac{\partial p}{\partial R}-\frac{\gamma^2 w_t^2v_\phi^2}{R}+\frac{b_\phi^2}{R}=0. \label{eq:equilibrium}
\end{equation}
The jets in this study do not rotate, so magnetic tension alone provides the inward force that balances the total pressure.
We may therefore approximate $p_a\lesssim\langle p_j\rangle$, which implies $\Lambda(p_j)\lesssim\Lambda(p_a)$ and connects the criterion to the ambient gas pressure $p_a$.
As the jet expands and the magnetic tension weakens, the jet and ambient pressures approach equilibrium.

The new stability parameter highlights its dependence on both the jet power and the ambient-pressure profile.
The importance of jet power has already been demonstrated by \cite{tchekhovskoy2016}.
In ideal MHD, $b_z/b_\phi\propto1/R_j$.
Consequently, if $p_a\propto z^{-b}$ and $R_j\propto z^c$, then $\Lambda\propto z^{b/2-c-1}$.
Suppressing the growth of the CD kink instability requires $b\ge3$ for a parabolic jet and $b\ge4$ for a conical jet.
Because the jet shape itself also depends on the ambient-pressure profile \citep[e.g.,][]{lyubarsky2009,komissarov2009}, the evolution of $\Lambda$ is controlled primarily by the ambient pressure.

Figure~\ref{fig:para} compares $\Lambda(p_a)$ and $\Lambda(p_j)$ along each jet with the simpler parameter $\Lambda_M=10\gamma R_j/z$ proposed by \cite{mizuno2009kink}.
Although $\Lambda_M$ tends to overestimate the jet stability, our new parameter $\Lambda(p_j)$ correctly predicts the behavior of every model.
The ambient-pressure estimate $\Lambda(p_a)$ follows a similar trend.
The orange shading on the right side of each panel marks the region dominated by internal energy.
In models {\tt 3D} and {\tt 3D0}, the stability parameter falls below unity only near the boundary of this region.
The jets are therefore formally unstable there, but the kink growth rate is small, consistent with the weak distortions in Figure~\ref{fig:3d}(g)-(i).
By contrast, the jets in models {\tt 3Dl} and {\tt 3Dt} remain magnetically dominated over much greater distances, allowing the kink instability to grow downstream.
The new stability parameter therefore reproduces the CD-kink behavior in all four simulations.

\subsection{Ray-traced Images of the Jets}\label{rad}

Having analyzed the jet dynamics, we now examine the radiative signatures of models {\tt 3D} and {\tt 3Dl}.
We perform SRRT calculations with {\tt RaptorP} \citep{hu2025apj}, a special-relativistic extension of the polarized general relativistic radiative transfer code {\tt RAPTOR} \citep{bronzwaer2018,bronzwaer2020}.
The calculations are performed in Minkowski spacetime under the fast-light approximation and include synchrotron emission and absorption as well as Faraday rotation.
{\tt RaptorP} directly supports the stretched grid used for the {\tt PLUTO} simulation data.

The SRRT calculations require a black hole mass and a conversion of the plasma density to cgs units.
We adopt parameters appropriate for M87: a black hole mass of $6.5\times10^9\,M_\odot$ and a distance of $16.8\,\mathrm{Mpc}$.
We set the jet power to $6.6\times10^{42}\,\mathrm{erg\,s^{-1}}$, which gives $M_{\rm unit}=1.2\times10^{20}\,\mathrm{g}$.
With these units, $z=6\times10^4$ corresponds to 18.7~pc, and the characteristic magnetic-field strength is approximately $10\,\mathrm{mG}$.

We use a $\kappa$ distribution to represent a hybrid thermal--nonthermal electron population \citep{xiao2006}:
\begin{equation}
  \frac{d N}{d\gamma_e} \propto \gamma_e\sqrt{\gamma_e^2-1}\left(1+\frac{\gamma_e-1}{\kappa w}\right)^{-(\kappa+1)}, \label{eq12}
\end{equation}
where $\gamma_e$ is electron Lorentz factor.
To determine $\kappa$ and $w$, we adopt the results of particle-in-cell simulations of decaying plasma turbulence by \cite{meringolo2023}, who parameterized the nonthermal energy production efficiency $\epsilon$ and power-law index $\kappa$ in turbulent plasmas as
\begin{align}
    \epsilon &=1-\frac{0.23}{\sqrt{\sigma}}+0.5\sigma^{0.1}\tanh(-10.18\sigma^{0.1}\beta),\\
    \kappa &=2.8+\frac{0.2}{\sqrt{\sigma}}+1.6\sigma^{-0.6}\tanh(2.25\beta\sigma^{1/3}). \label{eq:turbulence}
\end{align}
Here we assume electron and proton temperatures are equal and determine $w$ using the method described in Appendix~B of \cite{hu2025apj}.

We perform the SRRT calculations at $t=6\times10^4$, corresponding to approximately 60.9~yr in physical units. 
We include nearly the entire simulation domain.
The field of view is $1\times10^4$ by $5\times10^4$ in code units.
To compute the spectral energy distribution (SED), we sample frequencies from $0.1$ to $10^5$~GHz.
The viewing angle is fixed at $90^\circ$.

\begin{figure}[!h]
  \centering
  \includegraphics[width=\linewidth]{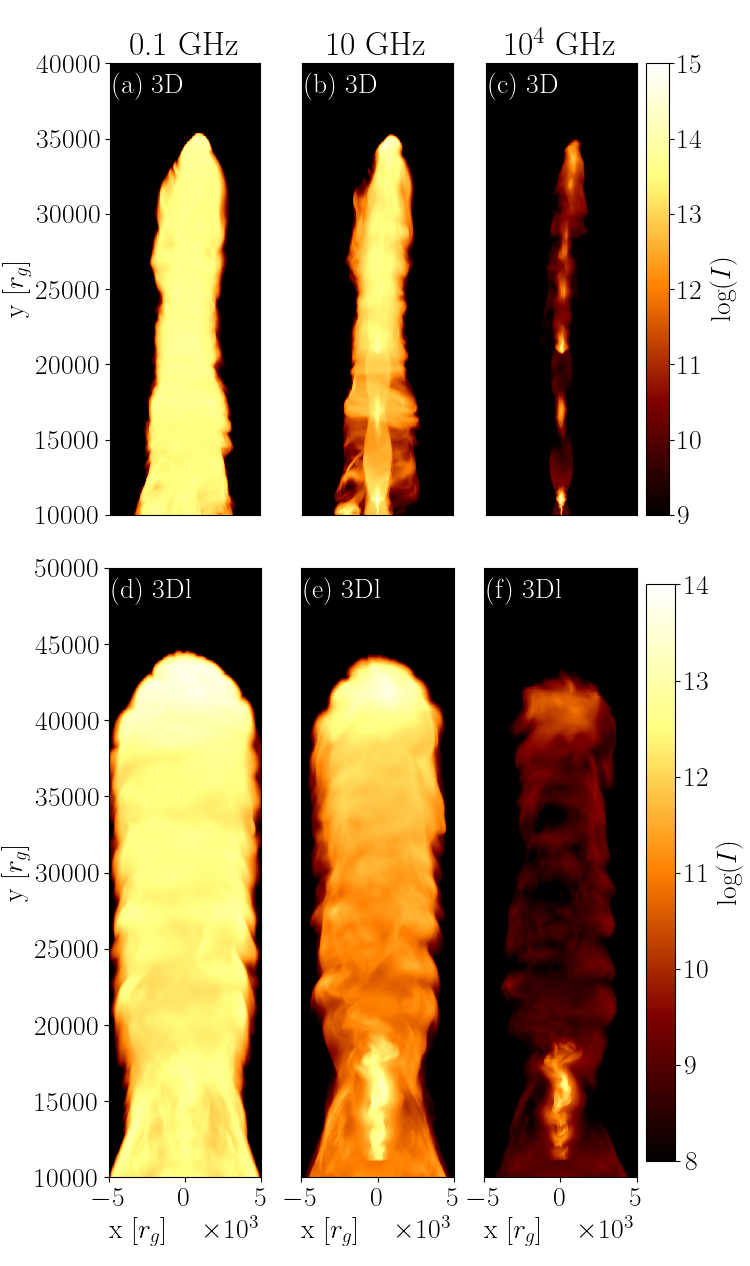}
  \caption{Multi-frequency ray-traced images of models {\tt 3D} (top row) and {\tt 3Dl} (bottom row) at $t=6\times10^4$. The viewing angle is $90^\circ$.}
  \label{fig:rad}
\end{figure}

Figure~\ref{fig:rad} shows ray-traced images of models {\tt 3D} and {\tt 3Dl} at 0.1, 100, and $10^4$~GHz.
At 0.1~GHz, both jets are optically thick, and the brightest emission originates from the termination shock at the jet head.
Their morphologies diverge at higher frequencies due to reduced optical thickness.
In model {\tt 3D}, the recollimation shocks appear as a series of bright knots, similar to the knot complex observed downstream of HST-1 \citep{thimmappa2024}.
In model {\tt 3Dl}, the high-frequency emission is likewise dominated by the recollimation shock.
Because the CD kink instability starts to grow at this location, the associated twisted structure is also clearly visible.

\section{Discussion and Conclusions}\label{con}

We have systematically investigated how the radial profile of the ambient medium affects the large-scale evolution of relativistic magnetized jets.
Our simulations span three orders of magnitude in distance.
The jets initially propagate through a Bondi-like atmosphere, converting magnetic energy into bulk kinetic energy.

The external pressure plays a central role in determining the jet geometry during this process.
Pressure-balanced models produce a broad parabolic jet, $R_j\propto z^{0.649\pm0.003}$, whereas lower ambient pressures produce a conical jet.
In M87, the observed transition occurs approximately $10^5$-$10^6\,r_g$ \citep[e.g.,][]{asada2012,hada2024}.
Maintaining a parabolic jet over such a large distance may be difficult in a steep atmosphere such as an ideal Bondi profile.
Therefore, additional confinement mechanisms or deviations from the Bondi profile may be required.
One possibility is a multi-component outflow consisting of a jet spine and a surrounding disk-wind sheath, as naturally produced in GRMHD simulations of jet formation \citep[e.g.,][]{ressler2020,cho2023,guo2025}.

At the downstream of the first recollimation shock, the jet evolution depends on the dominant components of energy.
The internal energy-dominated jets in models {\tt 3D} and {\tt 3D0} develop a series of recollimation shocks.
By contrast, the magnetically dominated jets in models {\tt 3Dl} and {\tt 3Dt} grow an internal CD kink instability.
Some observed AGN jets, including M87, have a parabolic shape near the black hole.
Their narrow cross sections favor strong recollimation, which may in turn promote the growth of the CD kink instability \citep{duran2017}.
The parabolic geometry, jet recollimation, and kink growth may therefore be physically connected.

To clarify the relationship between the ambient-medium profile and the growth of the CD kink instability, we derived a new stability parameter that depends on the jet power and ambient-pressure profile.
Its predictions are consistent with the behavior of all simulated models.

{Using physical parameters appropriate for M87, we post-processed the simulations with the SRRT code to calculate the emission.
At low frequencies, the jets are optically thick, and the emission is dominated by termination shocks at the jet head.
At higher frequencies, the recollimation shocks become exceptionally bright due to reduced optical thickness.
In model {\tt 3D}, they appear as a series of knots resembling the complex downstream of HST-1.
A quantitative comparison with observations will, however, require a more realistic ambient profile \citep[e.g.,][]{russell2015}.
In jets with a single prominent recollimation shock, the brightest component lies relatively far upstream, producing an FR~I-like emission morphology.
Growth of the kink instability is observed at higher frequencies.

A further limitation is that the injected jet is prescribed rather than generated self-consistently, and its initial magnetic configuration can affect the subsequent large-scale evolution \citep{hu2025aa}.
A fully self-consistent calculation would follow jet formation near the central black hole with GRMHD and then propagate the outflow to much larger scales, which remains computationally demanding.
To address this limitation, we are developing a framework that uses GRMHD-generated outflows as the injection conditions for large-scale SRMHD simulations.

\begin{acknowledgements}
We thank Jie-Shuang Wang for helpful advice.
We acknowledge the referee for valuable advice, which greatly improved the quality of the paper.
This work was supported by the National Key Research and Development Program of China (grant No. 2023YFE0101200), the National Natural Science Foundation of China (grant Nos. 12273022 and 12511540053), and the Shanghai Municipality Orientation Program of Basic Research for International Scientists (grant No. 22JC1410600).
The simulations were performed on the Astro cluster at the Tsung-Dao Lee Institute and the Siyuan-1 cluster at the Center for High Performance Computing, Shanghai Jiao Tong University.

\end{acknowledgements}

\appendix
\section{Derivation of the self-balanced magnetic profile}\label{mag}

The relativistic jet is usually considered magnetized and cold, which means the magnetic tension balances the gradient of magnetic pressure:

\begin{equation}
    \frac{d b^2/2}{dR}+\frac{b_\phi^2}{R}=0 \label{eq:bala}
\end{equation}

For the initial setup of jet simulation, we usually begin with $B_r=0,v_r=0,v_\phi\ll v_z$.
If we further assume a constant magnetic pitch $P=\frac{RB_z}{R_j B_\phi}$ across the jet, the equation~\ref{eq:bala} can be written as:

\begin{equation}
    -\frac{B_\phi(R)}{\gamma^2R^3}[(R^2-P^2R_j^2(1+v_z^2\gamma^2))B_\phi(R)+R(R^2+P^2R_j^2(1+vz^2\gamma^2))B_\phi'(R)]=0
    \label{eq:expand}
\end{equation}
which is a differential equation for $B_\phi(R)$. 
The general solution is $B_\phi(R)\propto R/(R^2+A)$, where $A=\gamma^2P^2R_j^2$.

\bibliography{main}{}
\bibliographystyle{aasjournalv7}
\end{document}